\documentstyle[11pt,epsfig]{article}
\title{\bf Effect of extrinsic curvature on quark--hadron phase transition}
\author{Malihe Heydari-Fard $^{2}$\thanks{email: heydarifard@qom.ac.ir}\hspace{0.5mm}
and Hamid Reza Sepangi $^{1}$\thanks{email: hr-sepangi@sbu.ac.ir}
\\ {$^{1}$ \small Department of Physics, Shahid
Beheshti University, G. C., Evin, Tehran 19839, Iran}\\{$^{2}$
\small Department of Physics, The University of Qom, Qom
37185-359, Iran} }
\begin{document}
\maketitle 
\begin{abstract}
The last phase transition predicted by the standard model of
particle physics took place at the QCD scale $T\sim200$ MeV when
the universe was about $t\sim10^{-5}$ seconds old and the Hubble
radius was around $10$ Km. In this paper, we consider the
quark--hadron phase transition in the context of brane-world
cosmology where our universe is a 3-brane embedded in a
$m$-dimensional bulk and localization of matter on the brane is
achieved by means of a confining potential. We study the behavior
of the physical quantities relevant to the description of the
early universe like the energy density, temperature and scale
factor, before, during, and after the phase transition and
investigate the effects of extrinsic curvature on the cosmological
phase transition. We show that the brane-world effects reduce the
effective temperature of the quark--gluon plasma and of the
hadronic fluid. Finally, we discuss the case where the universe
evolved through a mixed phase with a small initial supercooling
and monotonically growing hadronic bubbles.
\vspace{5mm}
\end{abstract}
\section{Introduction}
In the recent past, models with extra dimensions have been
proposed in which the standard fields are confined to a
four-dimensional ($4D$) world viewed as a hypersurface (the brane)
embedded in a higher dimensional space-time (the bulk) through
which only gravity can propagate. The most well-known model in the
context of brane-world theory is that proposed by Randall and
Sundrum (RS). In the so-called RSI model \cite{Randall1}, they
proposed a mechanism to solve the hierarchy problem with two
branes, while in the RSII model \cite{Randall2}, they considered a
single brane with a positive tension, where $4D$ Newtonian gravity
is recovered at low energies even if the extra dimension is not
compact. This mechanism provides an alternative to
compactification of extra dimensions. The cosmological evolution
of such a brane universe has been extensively investigated and
effects such as a quadratic density term in the Friedmann
equations have been found \cite{review,cosmology}. This term
arises from the imposition of the Israel junction conditions which
is a relationship between the extrinsic curvature and
energy-momentum tensor of the brane and results from the singular
behavior in the energy-momentum tensor. In brane theories the
covariant Einstein equations are also derived by projecting the
bulk equations onto the brane \cite{Shiromizu,Sh}. This was first
done by Shiromizu, Maeda, and Sasaki \cite{Shiromizu} where the
Gauss-Codazzi equations together with Israel junction conditions
were used to obtain the Einstein field equations on the 3-brane.
This method has predominantly been used in theories with one extra
dimension. If the number of extra dimensions exceeds one, no
reliable method for confining matter to the brane exists. This is
so since the requirement to define junction conditions is the
existence of a boundary (brane) which cannot be defined if the
number of extra dimensions is more than one. For example, a
boundary surface in a $3D$ space is a surface with one less
dimension whereas a line in the same space cannot be considered as
its boundary. On the back of such concerns, model theories have
been proposed where matter is confined to the brane through the
action of a confining potential, without the use of any junction
condition or $Z_2$ symmetry \cite{14}. In \cite{Maia,maia} the
authors used the confining potential approach to study a
brane-world embedded in a $m$-dimensional bulk. The field
equations so obtained on the brane contained an extra term which
was identified with the X-cold dark matter (XCDM). The dynamics of
test particles confined to a brane by the action of such a
potential at the classical and quantum levels were studied in
\cite{15}. In \cite{16}, a Friedmann-Robertson-Walker (FRW)
brane-world model was studied, offering a geometrical explanation
for the accelerated expansion of the universe. The same
methodology was used in \cite{20} to find the spherically
symmetric vacuum solutions of the field equations on the brane.
These solutions were shown to account for the accelerated
expansion of the universe and offered an explanation for the
galaxy rotation curves. The classical tests stemming from a
brane-world with a confining potential has been studied in
\cite{new}.

During the evolution of the very early universe there have been at
least two phase transitions. The electroweak theory predicts that
at about $100$ GeV there was a transition from a symmetric high
temperature phase with massless gauge bosons to the Higgs phase,
in which the $SU(2)\times U(1)$ gauge symmetry is spontaneously
broken and all the masses in the model are generated. A detailed
understanding of the nature and dynamics of this transition is a
very difficult task, and a lot of quantitative analytical and
numerical studies have been performed over the years. One of the
motivations of these studies has been that non-perturbative
effects might have changed the baryon asymmetry in the phase
transition. Also, the quantum chromodynamics (QCD) prediction that
at about $200$ MeV there was a transition from a quark--gluon
plasma to a plasma of light hadrons, is thought to have occurred
early in the history of the universe. The physical conditions for
this transition to take place, date it back to a few microseconds
after the Big Bang, when the universe had a mean density of the
same order as nuclear matter ($\rho\sim10^{15}$ g/cm$^{3}$). The
quark--hadron transition marks the end of the exotic physics of
the very early universe and the beginning of the era of processes
and phenomena which have a direct counterpart in the high energy
experiments now being carried out with modern accelerators. It is
also the last of the early universe phase transitions (at least
within the standard picture) and so could be relevant both as a
potential filter for the relics produced by previous transitions
and also as a best candidate for the production of inhomogeneities
which could have survived to later epochs \cite{lecture}.

As is well known, phase transitions are called first or second
order depending on whether the position of the vacuum state
changes discontinuously or continuously as the critical
temperature is reached. A first order quark--hadron phase
transition in the expanding universe can be described generically
as follows \cite{k}. The cooling down of the color deconfined
quark--gluon plasma below the critical temperature, believed to be
around $T_c\approx$150 MeV, makes it energetically favorable to
form color confined hadrons (mainly pions and a tiny amount of
neutrons and protons, since the net baryon number should be
conserved). However, the new phase does not show up immediately. A
first order phase transition generally needs some supercooling to
overcome the energy expended in forming the surface of the bubble
and the new hadron phase. When a hadron bubble is nucleated,
latent heat is released and a spherical shock wave expands into
the surrounding supercooled quark--gluon plasma. This causes the
plasma to reheat, approaching the critical temperature and
preventing further nucleation in a region passed by one or more
shock fronts. Bubble growth is generally described by
deflagrations where a shock front precedes the actual transition
front. The nucleation stops when the whole universe has reheated
$T_c$. The phase transition corresponding to this phase ends
promptly, in about 0.05 $\mu s$, during which the cosmic expansion
is completely negligible. Afterwards, the hadron bubbles grow at
the expense of the quark phase and eventually percolate or
coalesce. Ultimately, when all quark--gluon plasma has been
converted into hadrons, neglecting possible quark nugget
production, the transition ends. The physics of the quark--hadron
phase transition and its cosmological implications have been
extensively discussed in the framework of general relativistic
cosmology in \cite{phase}-\cite{transition}.

As was mentioned above, the Friedmann equation in brane-world
scenario contains deviations from $4D$ cosmology. We expect this
deviation from the standard $4D$ cosmology to have noticeable
effects on the cosmological evolution, especially on cosmological
phase transitions. In the context of brane-world models, the first
order phase transitions have been studied in \cite{Davis} where it
has been shown that due to the effects coming from higher
dimensions, a phase transition requires a higher nucleation rate
to complete and baryogenesis and particle abundances could be
suppressed. Recently, the quark--hadron phase transition has been
studied in a brane-world scenario by assuming that the phase
transition is of the first order. It has been shown that the
brane-world effects lead to an overall decrease of the temperature
of the very early universe, and accelerate the transition to the
pure hadronic phase \cite{Harko}. It would therefore be of
interest to study the latter in a brane-world model with a
constant curvature bulk without using the $Z_2$ symmetry and
without postulating any form of junction conditions
\cite{Maia,maia}. In so doing, the gravitational field equations
on the brane are modified by a local extra term, $Q_{\mu\nu}$.
Using the modified Friedmann equation and the equation of state of
the matter, we study the effects of extrinsic curvature on the
quark--hadron phase transition.

\section{The model}
Let us start by presenting the model used in our calculations. We
only state the results and refer the reader to \cite{maia,16} for
a detailed derivation of these results.

As was mentioned in the introduction, the brane-world model we
invoke here differs from the usual Randall-Sundrum type in that no
junction conditions or $Z_2$ symmetry is used. One thus starts
with the usual setup in which a $4D$ brane is embedded in a 5 or,
in general, $m$-dimensional bulk. Assuming that the bulk space has
constant curvature, one arrives at the following equations
\cite{maia}
\begin{eqnarray}
G_{\mu\nu} = \kappa_4^2T_{\mu\nu}-\Lambda g_{\mu\nu} +
Q_{\mu\nu}.\label{b4}
\end{eqnarray}
where $\Lambda$ is the effective cosmological constant in $4D$,
$T_{\mu\nu}$ is the energy-momentum tensor of the matter confined
to the brane and $Q_{\mu\nu}$ is a completely geometrical quantity
given by
\begin{eqnarray}\label{a19}
Q_{\mu\nu} = \frac{1}{\varepsilon}\left[
\left(K^{\rho}_{\,\,\,\mu}K_{\rho\nu}-KK_{\mu\nu}\right)-\frac{1}{2}
\left(K_{\alpha\beta }K^{\alpha\beta}-K^2\right)g_{\mu\nu}\right].
\end{eqnarray}
We note that $Q_{\mu\nu}$ is an independently conserved quantity,
that is
\begin{eqnarray}\label{a19}
Q^{\mu\nu}_{\,\,\,\ ;\mu}=0,
\end{eqnarray}
so that equation (\ref{b4}) satisfies the covariant conservation
law. Equation (\ref{b4}) is the starting point from which a class
of solutions were found in \cite{maia}. Thus, starting with the
metric
\begin{eqnarray}
ds^2 = -dt^2 + a(t)^2 \left[dr^2 + r^2 (d\theta^2 + \sin^2\theta
d\varphi^2)\right],\label{ds}
\end{eqnarray}
and calculating the components of extrinsic curvature from the
Codazzi equation, the gravitational field equations on the brane
become
\begin{equation}
\left(\frac{\dot{a}}{a}\right)^2 =
\frac{\kappa_4^2}{3}\rho+\frac{\Lambda}{3}+\frac{1}{\varepsilon}\frac{b^2}{a^4},\label{new}
\end{equation}
\begin{equation}
\frac{\ddot{a}}{a} =
-\frac{\kappa_4^2}{6}(\rho+3p)+\frac{\Lambda}{6}+
\frac{1}{\varepsilon}\frac{b^2}{a^4}\frac{1}{\dot{a}b}a^2\frac{d}{dt}\left(\frac{b}{a}\right),\label{New}
\end{equation}
\begin{equation}
\dot{\rho}+3\frac{\dot{a}}{a}(\rho+p)=0.\label{b22}
\end{equation}
Now, using the geometrical energy-momentum tensor for $Q_{\mu\nu}$
(XCDM), the Friedmann equations (\ref{new}) and (\ref{New}) can be
written as (for more details see \cite{maia})
\begin{equation}
\left(\frac{\dot{a}}{a}\right)^2 =
\frac{\kappa_4^2}{3}\rho+\frac{\Lambda}{3}+\frac{1}{\varepsilon}b_0^2
{a^{-3(1+w_x)}},\label{b21}
\end{equation}
\begin{equation}
\frac{\ddot{a}}{a} =
-\frac{\kappa_4^2}{6}(\rho+3p)+\frac{\Lambda}{6}-\frac{1}{2\varepsilon}{b_0^2}(1+3w_x)a^{-3(1+w_x)},\label{b}
\end{equation}
where $b_0$ is a constant of integration and $w_x$ is a constant
appearing in the equation of state, $p_x=w_x\rho_x$. As can be
seen, the correction term with respect to the standard Friedmann
equation is given by the components of the extrinsic curvature.

For the sake of completeness, let us compare the model presented
in this work to the usual brane-world models where the Israel
junction conditions is used to calculate the extrinsic curvature
in terms of the energy-momentum tensor and its trace on the brane.
If we did that, we would obtain
\begin{equation}
b(t) = -\frac{1}{6}\kappa_5^2\rho a^2,\label{b23}
\end{equation}
which, upon its substitution in equation (\ref{new}), gives
\begin{equation}
\left(\frac{\dot{a}}{a}\right)^2 =
\frac{\kappa_4^2}{3}\rho+\frac{\Lambda}{3}+\frac{1}{3}\frac{\kappa_4^2}{2\lambda}\rho^2,\label{b24}
\end{equation}
which is the same as equation (6) in \cite{Harko}. On the other
hand, for the perfect fluid with energy density $\rho = \rho_0
a^{-3(1+w)}$, equation (\ref{b21}) reduces to the Friedmann
equation (\ref{b24}) in the usual brane-world models if we take
$b_0=-\frac{1}{6}\kappa_5^2\rho_0$ and $w_x=1+2w$ .

\section{The quark--hadron phase transition}
We are going to consider phase transition in the context of the
brane-world model without mirror symmetry or any form of junction
condition. In a cosmological setting, these could be modified
because of the modified Friedmann equations. Let us first outline
the relevant physical quantities of the quark--hadron phase
transition which will be used in the following sections. In order
to study the quark--hadron phase transition we should specify the
equation of state of the matter, in both quark and hadron states.
In this regard, a well written and concise review can be found in
\cite{Harko} and the interested reader should consult it. Here, it
would suffice to mention the results relevant to our study and
leave the details of the discussion to the said reference.

For a first order phase transition in the quark phase the equation
of state of the matter can generally be given by
\begin{eqnarray}
\rho_q =
g^{*}_{q}\left(\frac{\pi^2}{30}\right)T^4+V(T),\hspace{0.5 cm}p_q
= g^{*}_{q}\left(\frac{\pi^2}{90}\right)T^4-V(T),\label{c1}
\end{eqnarray}
where $g^{*}_{q}=16+21/2(N_F=2)+14.25=51.25$ is the effective
number of degrees of freedom in the quark phase. We choose the
following expression for the self-interaction potential $V(T)$
\cite{VT}
\begin{eqnarray}
V(T) = B+\gamma_{T}T^2-\alpha_{T}T^4,\label{c2}
\end{eqnarray}
where $B$ is the bag pressure constant, $\alpha_T = {7\pi^2}/{20}$
and $\gamma_T = {m^2_s}/{4}$ with $m_s$ being the mass of the
strange quark in the range $m_s\in(60-200)$ MeV. The form of the
potential $V$ corresponds to a physical model in which the quark
fields are interacting with a chiral field formed with the $\pi$
meson field and a scalar field. By ignoring the temperature
effects, the equation of state in the quark phase takes the form
of the MIT bag model equation of state, $p_q = (\rho_q -4B)/3$.
Results obtained in low energy hadron spectroscopy, heavy ion
collisions and phenomenological fits of light hadron properties
give a range for $B^{1/4}$ between $100-200$ MeV \cite{MIT}.

For the hadron phase we take the cosmological fluid with energy
density $\rho_h$ and pressure $p_h$ as an ideal gas of massless
pions and of nucleons described by the Maxwell-Boltzmann
statistics. The equation of state can be approximated by
\begin{eqnarray}
p_h(T) = \frac{1}{3}\rho_{h}(T) = g^{*}_{h}
\left(\frac{\pi^2}{90}\right)T^4,\label{c4}
\end{eqnarray}
where $g^{*}_h=17.25$.

During the quark--hadron phase transition the temperature is equal
to the critical temperature $T_c$ which is defined by the
condition $p_q(T_c) = p_h(T_c)$ \cite{k}, and is given by
\begin{eqnarray}
T_c =
\left[{\frac{\gamma_T+\sqrt{\gamma_T^2+4B\left(\alpha_T+a_q-a_\pi\right)}}
{2\left(\alpha_T+a_q-a_\pi\right)}}\right]^{1/2},\label{c3}
\end{eqnarray}
where $a_q=(\pi^2/90)g^*_q$ and $a_\pi=(\pi^2/90)g^*_h$. For $m_s
= 200$ MeV and $B^{1/4} = 200$ MeV the transition temperature is
of the order $T_c\approx125$ MeV. It is worth mentioning that
since the phase transition is assumed to be of first order, all
the physical quantities, like the energy density, pressure and
entropy exhibit discontinuities across the critical curve. In the
next section, we consider the evolution of the brane-world
scenario before, during and after the phase transition era.

\section{Dynamical evolution during the phase transition}
To begin with, we obtain the physically important quantities in
the quark--hadron phase transition in the brane-world scenario,
like the energy density $\rho$, temperature $T$ and  scale factor
$a$. These parameters are derived from the modified Friedmann
equation (\ref{b21}), conservation equation (\ref{b22}) and the
equations of state (\ref{c1}), (\ref{c2}) and (\ref{c4}).

$\bullet$ For $T > T_c$, before the phase transition, the
brane-world is in the quark phase. Use of the equations of state
of the quark matter and conservation of matter on the brane, we
can rewrite equation (\ref{b22}) as
\begin{eqnarray}
\frac{\dot{a}}{a} =
-\frac{3a_q-\alpha_T}{3a_q}\frac{\dot{T}}{T}-\frac{\gamma_T}{6a_q}\frac{\dot
{T}}{T^3},\label{d1}
\end{eqnarray}
the general solution of equation (\ref{d1}) becomes
\begin{eqnarray}
a(T) =
a_0T^{\frac{\alpha_T-3a_q}{3a_q}}\exp\left(\frac{\gamma_T}{12a_q}\frac{1}{T^2}\right),\label{d2}
\end{eqnarray}
where $a_0$ is a constant of integration.

Substituting equation (\ref{d1}) into modified Friedmann equation
(\ref{new}), one obtain
\begin{eqnarray}
\frac{dT}{dt} =
-\frac{T^3}{\sqrt{3}}\frac{\sqrt{\kappa_4^2(3a_q-\alpha_{T})T^4+\kappa_4^2\gamma_{T}T^2+\kappa_4^2
B+\Lambda+\frac{3}{\varepsilon}\frac{b^2}{a^4}}}
{(1-\frac{\alpha_T}{3a_q})T^2+ \frac{\gamma_T}{6a_q}}.\label{d3}
\end{eqnarray}
Now, use of equations (\ref{b23}) and (\ref{c1}) in above equation
lead to the variation of the temperature of the brane universe in
the standard brane-world models, which is same as equation (17) in
\cite{Harko} in the absence of nonlocal bulk effects.

Substituting equation (\ref{d1}) in modified Friedmann equation
(\ref{b21}), the basic equation describing the evolution of
temperature of the brane universe in the quark phase can be
written as
\begin{eqnarray}
\frac{dT}{dt} =
-\frac{T^3}{\sqrt{3}}\frac{\sqrt{(\Lambda+\kappa_4^2B)+\kappa_4^2\gamma_T
T^2+\kappa_4^2(3a_q-\alpha_T)T^4+\frac{3b_0^2}{\varepsilon}
\left[T^{-A_0}\exp\left({\frac{A_1}{2T^2}}\right)\right]^{-3(1+w_{x})}}}
{A_0T^2+A_1},\label{d3}
\end{eqnarray}
where we have denoted
$$
A_0 = 1-\frac{\alpha_T}{3a_q},
$$
$$
A_1 = \frac{\gamma_T}{6a_q}.
$$
The behavior of the temperature as a function of time in the quark
matter filled brane-world is illustrated, for $w_x=-0.5$ and
different values of the constant $b_0$, in figure 1. The behavior
of this parameter shows that the effects of extrinsic curvature
would significantly reduce the temperature of the quark--gluon
plasma and accelerate the phase transition to the hadronic era.
Here, the solid curve corresponds to the general relativistic
limit. In figure 2, we have illustrated the behavior of the
temperature of the brane in the quark phase for $b_0^2=10^8$ and
different values of $w_x$. In reference \cite{16}, we have
discussed the role of extrinsic curvature in dark energy. We have
found that the accelerating expansion of the universe can be a
consequence of the extrinsic curvature and thus a purely
geometrical effect. The rate of expansion and so the age of the
universe increase with decreasing $w_x$. Here, as one can see from
figure 2, the temperature evolution of the early universe is also
sensitive to this parameter and increases with decreasing $w_x$.

One can obtain an analytical insight into the evolution of the
cosmological quark matter in  brane-world scenarios by looking at
the effects of the extrinsic curvature on the phase transition.
Let us consider the case in which temperature corrections can be
neglected in the self-interacting potential $V$, that is
$V=B=const.$ with the equation of state of the quark matter being
given by the bag model, $p_q=(\rho_q-4B)/3$. Thus, equation
(\ref{b22}) leads to the following scale factor
\begin{eqnarray}
a(T) = \frac{a_0}{T}, \label{D4}
\end{eqnarray}
where $a_0$ is a constant of integration. Taking $w_x>1/3$ in
equation (\ref{b21}), the correction term due to the extrinsic
curvature ($\propto a^{-3(1+w_x)}\propto T^{3(1+w_x)}$) is the
dominant term relative to the energy density of the matter in the
quark phase $(\propto T^4)$. Therefore, the evolution of the quark
phase of the brane-world is now described by the equation
\begin{eqnarray}
\frac{\dot{a}}{a}\approx \frac{b_0}{a_0^{3/2(1+w_x)}}
T^{3/2(1+w_x)}. \label{D5}
\end{eqnarray}
Now, use of equations (\ref{D4}) and (\ref{D5}) lead to the
following equation for the temperature
\begin{eqnarray}
\frac{dT}{dt}\approx-\frac{b_0}{a_0^{3/2(1+w_x)}}T^{1/2(5+3w_x)},\label{D6}
\end{eqnarray}
with the general solution given by
\begin{eqnarray}
T(t)\approx\left[\frac{3}{2}b_0a_0^{-3/2(1+w_x)}(1+w_x)t\right]^{-\frac{2}{3(1+w_x)}}.\label{D7}
\end{eqnarray}

\begin{figure}
\centerline{\begin{tabular}{ccc}
\epsfig{figure=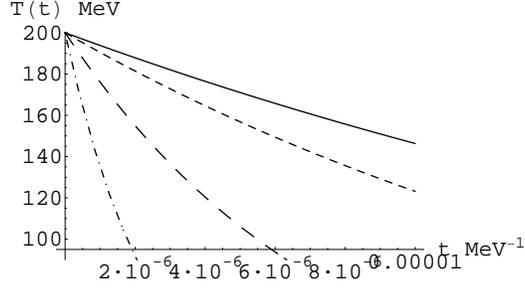,width=7cm}
\end{tabular} } \caption{\footnotesize The behavior of temperature of the quark
fluid on the brane as a function of time for different values of
$b_0^2$, $0$ (solid curve), $10^9$ (dotted curve), $10^{10}$
(dashed curve) and $10^{11}$ (dotted-dashed curve) with $B^{1/4} =
200$ MeV, $w_x=-0.5$ and $\varepsilon=+1$.}\label{figure1}
\end{figure}

\begin{figure}
\centerline{\begin{tabular}{ccc} \epsfig{figure=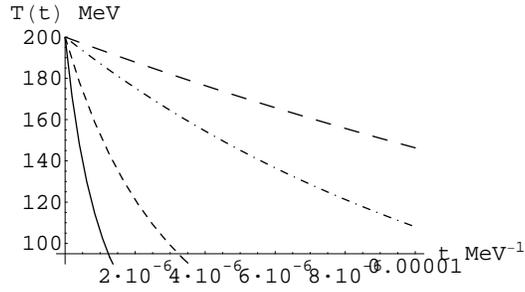,width=7cm}
\end{tabular} } \caption{\footnotesize The behavior of temperature of the quark
fluid on the brane as a function of time for different values of
the $w_x$, $-1/3$ (solid curve), $-0.5$ (dotted curve), $-0.75$
(dotted-dashed curve) and $-1$ (dashed curve) with $B^{1/4} = 200$
MeV, $b_0^2=$$10^8$ and $\varepsilon=+1$.}\label{figure1}
\end{figure}

$\bullet$ For ${T = T_c}$, during the phase transition, the
temperature and the pressure are constant. In this case the
entropy $S = sa^3$ and the enthalpy $W = (\rho + p)a^3$ are
conserved. For later convenience, following \cite{k}, we replace
$\rho(t)$ by $h(t)$, so that the volume fraction of matter in the
hadron phase is given by
\begin{eqnarray}
\rho(t) = \rho_H h(t) + \rho_Q [1-h(t)] = \rho_Q[1+
nh(t)],\label{d4}
\end{eqnarray}
where $n = (\rho_H-\rho_Q)/\rho_Q$. At the beginning of the phase
transition $h(t_c) = 0$, where $t_c$ is the time representing the
beginning of the phase transition and $\rho(t_c)\equiv \rho_Q$,
while at the end of the transition $h(t_h) = 1$, where $t_h$ is
the time at which the phase transition ends, representing
$\rho(t_h)\equiv\rho_H$. For $t > t_h$ the universe enters the
hadronic phase. Substituting equation (\ref{d4}) into equation
(\ref{b22}) we have
\begin{eqnarray}
\frac{\dot{a}}{a} = -\frac{1}{ 3} \frac{(\rho_H-\rho_Q){\dot {h}}
}{ [\rho_Q +p_c +(\rho_H-\rho_Q)h]} .\label{d5}
\end{eqnarray}
The above equation leads to the following scale factor
\begin{eqnarray}
a(t) = a(t_c) \left[1+\frac{\rho_H-\rho_Q}{\rho_Q
+p_c}h(t)\right]^{-1/3},\label{d6}
\end{eqnarray}
where the initial condition $h(t_c) = 0$ has been used.

Substituting equation (\ref{d5}) into modified Friedmann equation
(\ref{new}), one obtains
\begin{eqnarray}
\frac{dh}{dt} = -\sqrt{3}\left(h+\frac{\rho_Q
+p_c}{\rho_H-\rho_Q}\right){\sqrt{\kappa_4^2\rho_Q[1+nh]+
\Lambda+\frac{3}{\varepsilon}\frac{b^2}{a^4}}}.\label{d3}
\end{eqnarray}
Now, use of equations (\ref{b23}) and (\ref{d4}) in the above
equation leads to the variation of temperature of the brane
universe in the standard brane-world models, which is the same as
equation (29) in \cite{Harko} in the absence of nonlocal bulk
effects.

The evolution of the fraction of the matter for the hadron phase
in our model is given by
\begin{eqnarray}
\frac{dh}{dt} = -\sqrt{3}\left(h+\frac{\rho_Q
+p_c}{\rho_H-\rho_Q}\right)\sqrt{\kappa_4^2\rho_Q[1+nh]+\Lambda+\frac{3b_0^2}{\varepsilon}
\left[{a(t_c)}\left(1+ \frac{\rho_H-\rho_Q}{\rho_Q
+p_c}h\right)^{-1/3}\right]^{-3(1+w_x)}}.\label{d7}
\end{eqnarray}

In figure 3, the variation of the hadron fraction as a function of
time is presented, for $w_x=-0.5$ and different values of $b_0$.
The behavior of this parameter shows that the hadron fraction is
again strongly dependent on the brane effects so that in the
presence of the term $Q_{\mu\nu}$ , $h(t)$ is much larger than the
standard general relativity. Figure 4 shows variation of the
hadron fraction as a function of time, for $b_0^2=10^8$ and
different values of $w_x$.

\begin{figure}
\centerline{\begin{tabular}{ccc}
\epsfig{figure=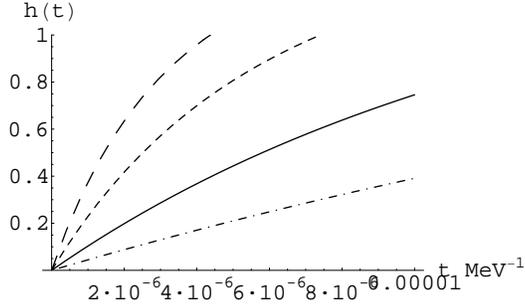,width=7cm}
\end{tabular} } \caption{\footnotesize The behavior of hadron fraction $h$
as a function of time during the quark-hadron phase transition on
the brane for different values of the $b_0^2$, $0$ (solid curve),
$10^9$ (dotted-dashed curve), $10^{10}$ (dashed curve) and
$10^{11}$ (dotted curve) with $B^{1/4} = 200$ MeV, $w_x=-0.5$ and
$\varepsilon=+1$.}\label{figure1}
\end{figure}

\begin{figure}
\centerline{\begin{tabular}{ccc} \epsfig{figure=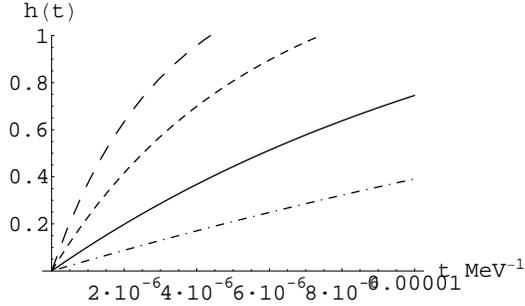,width=7cm}
\end{tabular} } \caption{\footnotesize The behavior of hadron fraction $h$
as a function of time during the quark-hadron phase transition on
the brane for different values of the $w_x$, $-0.75$ (solid
curve), $-0.5$ (dotted curve), $-1$ (dotted-dashed curve) and
$-1/3$ (dashed curve) with $B^{1/4} = 200$ MeV, $b_0^2=$$10^8$ and
$\varepsilon=+1$.}\label{figure1}
\end{figure}

$\bullet$ For $T < T_c$, after the phase transition, the energy
density of the pure hadronic matter is $\rho_h = 3p_h =
3a_{\pi}T^4$. The conservation equation on the brane (\ref{b22})
gives $a(T ) = a(t_h)T_c/T$ . The time variation of the
temperature of the brane universe in the hadronic phase is given
by
\begin{eqnarray}
\frac{dT}{dt}
=-\frac{T}{\sqrt{3}}\sqrt{\Lambda+3\kappa_4^2a_{\pi}T^4+
\frac{3b_0^2}{\varepsilon}\left[{a(t_h)T_c}\right]^{-3(1+w_x)}T^{3(1+w_x)}}.\label{d8}
\end{eqnarray}
The behavior of temperature of the hadron fluid filled brane
universe as a function of time for $w_x=-0.5$ and different values
of $b_0$ is presented in figure 5. As mentioned before the
variation of the temperature of the brane universe is sensitive to
both of parameters $w_x$ and $b_0$. Since $a(t_h)$ is a small
value quantity, we can obtain interesting effects on the
temperature of the very early universe for the small extrinsic
curvature if we take $w_x>-1$ and $a(t_h)=0.1$. Thus, the behavior
of the temperature as a function of time for $w_x=3/2$ and
different values of $b_0$ is represented in figure 6. The behavior
of the temperature for the brane in the hadron phase for
$b_0^2=10^8$ and different values of $w_x$ is presented in figure
7.

It is interesting to note that  for $w_x=\frac{5}{3}$,
$b_0=-\frac{1}{2}\kappa_5^2a_{\pi}a^4(t_h)T_c^4$ and
$\kappa_4^2=\frac{\lambda\kappa_5^4}{6}$, equation (\ref{d8})
reduces to equation (34) in \cite{Harko} where the variation of
temperature of the brane universe in the hadronic phase has been
studied within the context of standard brane-world models.

\begin{figure}
\centerline{\begin{tabular}{ccc}
\epsfig{figure=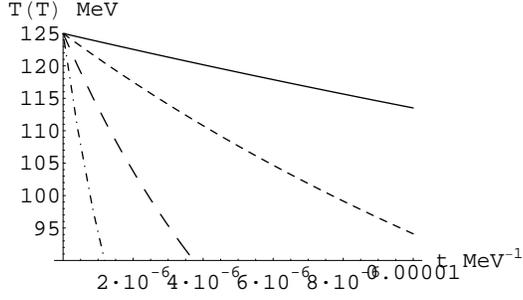,width=7cm}
\end{tabular} } \caption{\footnotesize The behavior of temperature of the
hadron fluid on the brane as a function of time for different
values of $b_0^2$, $0$ (solid curve), $10^9$ (dotted curve),
$10^{10}$ (dashed curve) and $10^{11}$  (dotted-dashed curve) with
$B^{1/4} = 200$ MeV, $a(t_h)=1$, $w_x=-0.5$ and
$\varepsilon=+1$.}\label{figure1}
\end{figure}

\begin{figure}
\centerline{\begin{tabular}{ccc} \epsfig{figure=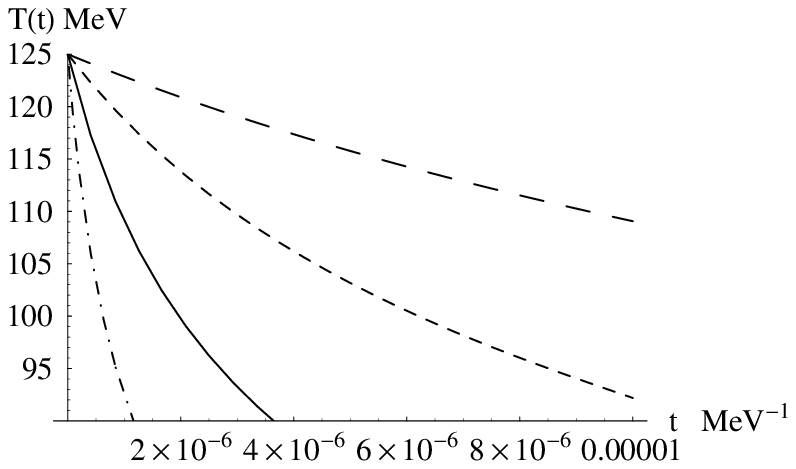,width=7cm}
\end{tabular} } \caption{\footnotesize The behavior of temperature of the
hadron fluid on the brane as a function of time for different
values of $b_0^2$, $10^3$ (solid curve), $10^2$ (dotted curve),
$10^{1}$ (dashed curve) and $10^4$  (dotted-dashed curve) with
$B^{1/4} = 200$ MeV, $a(t_h)=0.1$, $w_x=3/2$ and
$\varepsilon=+1$.}\label{figure1}
\end{figure}

\begin{figure}
\centerline{\begin{tabular}{ccc} \epsfig{figure=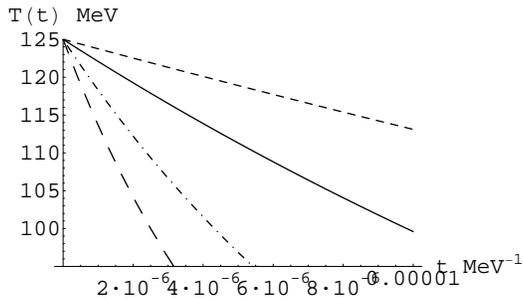,width=7cm}
\end{tabular} } \caption{\footnotesize The behavior of temperature of the
hadron fluid on the brane as a function of time for different
values of  $w_x$, $-1/3$ (dashed curve), $-0.75$ (solid curve),
$-1$ (dotted curve) and $-0.5$ (dotted-dashed curve) with $B^{1/4}
= 200$ MeV, $a(t_h)=1$, $b_0^2=$$10^8$ and
$\varepsilon=+1$.}\label{figure1}
\end{figure}

\section{Bubble nucleation in brane cosmology}
In this section, we consider the process of the formation and
evolution of microscopic quark nuclei in the cosmological fluid on
the brane using nucleation theory. The nucleation theory computes
the probability that a bubble or droplet of the A phase appears in
a system initially in the B phase near the critical temperature
\cite{Landau}.

In the previous section we discussed the mechanisms of  phase
transition and for the sake of completeness, we consider the phase
transition in the context of bubble nucleation theory in this
section. Phase transition can be described by an effective
nucleation theory in the small supercooling regime. As the
temperature decreases, there is a probability that a droplet of
hadrons is nucleated from the quark plasma. The nucleation
probability density is given by \cite{k}
\begin{eqnarray}
p(t) = p_0T_c^4\exp\left[-\frac{w_0}{(1-{\hat
T}^4)^2}\right],\label{B1}
\end{eqnarray}
where $p_0$ , ${w}_0$ are dimensionless constants and $\hat{T} =
T/Tc$, where $T_c$ is a critical temperature of the phase
transition. When a droplet of hadrons is formed, it starts to
expand explosively \cite{Gy}, with a velocity $v_{sh}$ smaller
than the sound speed. At the same time, a much quicker shock wave
is generated. During the period over which the temperature
decreases, a large number of bubbles are created until the shock
waves collide and re-heat the plasma to the critical temperature.
The fraction of the volume which at a time $t$ is turned into the
hadronic phase in the small supercooling scenario is given by
\cite{Harko}
\begin{eqnarray}
f(t) =
\int_{t_c}^{t}dt'p(t')\frac{4\pi}{3}v_{sh}^3(t-t')^3,\label{B2}
\end{eqnarray}
where $t_c$ is the time at which the critical temperature is
reached, and the nucleation process started.

It is now appropriate to consider the effects of the extrinsic
curvature in bubble nucleation in the early universe. Thus, using
equations (\ref{D5}) and (\ref{D6}), we can express this integral in
terms of the normalized temperature. The integral (\ref{B2})  may
now be given by
\begin{eqnarray}
f(\hat{T}) = \int_{\hat{T}}^{1}
\frac{d\hat{T'}}{\hat{T'}}\frac{p_0T_c^4v_{sh}^3}{b_0^4}\left(\frac{\hat{T'}}{c}\right)
^{-3/2(1+w_x)}\exp\left[-\frac{\omega_0}{(1-\hat{T'}^4)^2}\right]
\left[\frac{\hat{T}^{-3/2(1+w_x)}-\hat{T'}^{-3/2(1+w_x)}}
{3/2(1+w_x)c^{-3/2(1+w_x)}}\right]^3,\label{B3}
\end{eqnarray}
where $c=a_0/T_c$. To study the cosmological dynamics we use
numerical methods for calculating the integral. We have plotted
$f(\hat{T})$ against $\hat{T}$ for different values of $b_0^2$ and
$w_x=0.5$ with $v_{sh} = 10^{-3}$ and $w_0 = p_0$ = 1, in figure
8. The fraction of hadronic matter stays very close to zero until
the supercooling temperature is between $\hat{T} = 0.976$ and
$\hat{T} = 0.979$, then it jumps to 1 very rapidly. Here we may
deduce the result that at a certain temperature below the critical
value, an enormous amount of hadronic bubbles are nucleated almost
everywhere in the plasma, which grow explosively to transform all
the plasma into hadrons. This result is similar to what happens in
standard cosmology \cite{k} where the small supercooling phase
transition is also very rapid with respect to the simple first
order phase transition, at a temperature ($T\approx0.979$), which
is very similar to the one we have obtained in the context of
brane-world scenario.

\begin{figure}
\centerline{\begin{tabular}{ccc}
\epsfig{figure=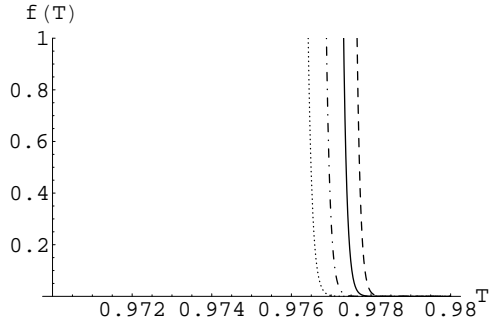,width=7cm}
\end{tabular} } \caption{\footnotesize The behavior of $f(\hat{T})$ as
a function of the normalized temperature for different values of
$b_0^2$, $10^8$ (dashed curve), $10^9$ (solid curve), $10^{10}$
(dotted-dashed curve) and $10^{11}$ (dotted curve) with $B^{1/4} =
200$ MeV, $w_x=0.5$ and $\varepsilon=+1$.} \label{figure1}
\end{figure}

\section{Conclusions}
The Friedmann equation in a brane-world scenario deviates from
that of the standard $4D$ case which imposes fundamental
phenomenological consequences on the cosmological evolution, and
in particular on the cosmological phase transitions. It has been
found that due to the effects coming from higher dimensions, the
temperature evolution of the universe in the brane-world scenario
is different from the standard FRW model \cite{Harko}.

In the present paper, we have investigated the quark--hadron phase
transition in a brane-world scenario in which the localization of
matter on the brane is achieved through the action of a confining
potential. We have studied the evolution of the physical
quantities relevant to the physical description of the early
universe like the energy density, temperature and scale factor,
before, during, and after the phase transition. We have shown that
for different values of parameters phase transition occurs and
results in decreasing the effective temperature of the
quark--gluon plasma and of the hadronic fluid. Bubble nucleation
which is one of different simplified mechanisms used to describe
the dynamics of a first order phase transition wa also studied.

\end{document}